\documentclass[fleqn,usenatbib]{mnras}

\usepackage{newtxtext,newtxmath}
\usepackage[T1]{fontenc}

\DeclareRobustCommand{\VAN}[3]{#2}
\let\VANthebibliography\thebibliography
\def\thebibliography{\DeclareRobustCommand{\VAN}[3]{##3}\VANthebibliography}

\usepackage{graphicx}	% Including figure files
\usepackage{amsmath}	% Advanced maths commands

\newcommand{\chandra}{\textit{Chandra}}

\newcommand{\xmm}{\textit{XMM-Newton}}

\newcommand{\nustar}{\textit{NuSTAR}}

\newcommand{\nicer}{\textit{NICER}}
\newcommand{\ms}{\ensuremath{M_{\odot}}}

\newcommand{\fluxcgs}{\ensuremath{\mathrm{erg}\,\mathrm{s}^{-1}\,\mathrm{cm}^{-2}}}
\newcommand{\lumcgs}{\ensuremath{\mathrm{erg}\,\mathrm{s}^{-1}}}

\title[Spectral variability in NGC 1042 ULX1]{Spectral variability in NGC 1042 ULX1}

\author[T. Ghosh and V. Rana]{
Tanuman Ghosh,\thanks{E-mail: tanuman@rri.res.in}
Vikram Rana
\\
Astronomy and Astrophysics, Raman Research Institute, C. V. Raman Avenue, Sadashivanagar, Bangalore 560080, India
}

\date{Accepted XXX. Received YYY; in original form ZZZ}

\pubyear{2022}

\begin{document}
\label{firstpage}
\pagerange{\pageref{firstpage}--\pageref{lastpage}}
\maketitle

\begin{abstract}
We report X-ray spectral variability in an ultraluminous X-ray source NGC 1042 ULX1, using archival \xmm\ and recent \nustar\ observations. In long-term evolution, the source has shown a trend of variation in spectral hardness. The variability in different \xmm\ observations is prominent above $\sim 1$ keV. Cool thermal disk component with a characteristic temperature of $\sim 0.2$ keV manifests that the spectral state of NGC 1042 ULX1 in all epochs is similar to that of the ultraluminous state sources. An apparent anti-correlation between luminosity and powerlaw index demonstrates that the source becomes spectrally harder when it is in a brighter state. That is conceivably related to variation in accretion rate, strength of comptonization, wind/outflow in the system or a manifestation of varying disk occultation. Typical hard ultraluminous type spectra indicate that NGC 1042 ULX1 is a low inclination system in general. Spectral properties suggest that, like many other ULXs which show spectral curvature around $\sim 6-10$ keV, NGC 1042 ULX1 could be another stellar-mass super-Eddington accretor.
\end{abstract}


%%%%%%%%%%%%%%%%%%%%%%%%%%%%%%%%%%%%%%%%%%%%%%%%%%

%%%%%%%%%%%%%%%%% BODY OF PAPER %%%%%%%%%%%%%%%%%%

\section{Introduction}

Ultraluminous X-ray sources (ULXs) are point-like off nuclear sources, X-ray luminosity of which is greater than the Eddington limit of a $10$ \ms\ black hole; $ \rm L_X > 10^{39}$ \lumcgs (see \citealt{Kaaret2017} for a recent review). Initially, these sources were considered to be the population of intermediate-mass black holes (IMBHs) accreting in sub-Eddington accretion rate \citep{Colbert1999}. However, unique spectral properties like curvature $\lesssim 10$ keV and presence of soft excess $\lesssim 0.4$ keV, characterize these sources as distinct from the sub-Eddington sources (e.g., \citealt{Stobbart2006, Gladstone2009, Sutton2013}). Especially, recent studies with broadband X-ray data strongly indicate that most of these sources are stellar-mass super-Eddington accretors (e.g., \citealt{Bachetti2013, Walton2013, Walton2014, Walton2015, Walton2015J, Mukherjee2015, Rana2015}). The discovery of a few neutron stars in these extra-galactic ULX populations further confirmed that super-Eddington accretion is a feasible scenario in some ULXs \citep{Bachetti2014, Furst2016N, Israel2017F, Israel2017M, Brightman2018, Carpano2018, Rodriguez2020, Sathyaprakash2019}. However, some ULXs have luminosities more than $10^{41}$ \lumcgs. They are known as hyperluminous X-ray sources (HLXs) and possible candidates for IMBH hosts (e.g., \citealt{Brightman2016, Webb2010}).

ULXs are broadly classified into different states based on spectral components and hardness (see \citealt{Kaaret2017, Sutton2013}). ULXs with a cool accretion disk and a powerlaw tail are classified as the ultraluminous state. Depending on the hardness of the source, they can be hard ultraluminous (HUL) or soft ultraluminous (SUL) regimes, as defined by \citealt{Sutton2013}. A correlated classification of these two regimes is hard intermediate and soft bright states, respectively, as defined in \citealt{Gurpide2021a, Gurpide2021b}. Typically, soft bright states are found in a higher luminosity regime than the hard intermediate state, which directly relates to the physical scenario of photons being down scattered by the dense medium of clumpy winds when the rate of accretion flow is high. The other states are broadened disk (BD) state and the super-soft ultraluminous (SSUL) state. The spectra of the former are dominated by thermal emission from a geometrically modified accretion disk. The SSUL spectra on the other hand, are dominated by single component cool blackbody emission. However, it is essential to understand that the distinction between these classifications is not always austere, and many ULXs show spectral transition among these different states/regimes (e.g., \citealt{Sutton2013, Walton2020, Gurpide2021a, Gurpide2021b, Dai2021}). These changes are due to variations in accretion rate, disk occultation, wind outflow strength, and several other physical parameters.

\citet{Kajava2009} studied a sample of ULXs which showed distinctive luminosity-spectral photon index ($\rm L_{X}-\Gamma$) correlation and anti-correlation. Such correlation studies indicate dissimilitude in ULX accretion scenarios in different epochs. Nevertheless, some important factors need to be considered before concluding such correlations. The absorption component and soft continuum like cool accretion disk can degenerate with a powerlaw model (e.g., \citealt{Feng2009, Kajava2009, Pinto2017}) since the powerlaw component extends to the low energy without any bound. Thus, luminosity from the soft spectral counterparts is often misinterpreted. Hence, to understand whether such correlations come from real physical property, it is crucial to investigate and mitigate these ``artefacts''.

NGC 1042 ULX1 (2XMM J024025.6-082428) is an extreme ULX, peak X-ray luminosity of which reaches $\rm L_X \sim 5 \times 10^{40}$ \lumcgs \citep{Sutton2012}. The host galaxy is a SAB(rs)cd type galaxy with a distance of $\sim 18.9$ Mpc. \citet{Sutton2012} studied a sample of extreme ULXs using X-ray data from \xmm\ and \chandra\ observatories. NGC 1042 ULX1 is one of the extremely luminous ULX in that sample. We study ULX1 in this work utilizing seven \xmm\ observations and one \nustar\ observation. Three of these seven \xmm\ observations (0093630101, 0306230101, 0553300401) were studied by \citet{Sutton2012} in detail. We primarily focus on the spectral properties and variability of ULX1 in different \xmm\ observations and how different spectral parameters vary during these epochs. The data from \nustar\ indicate a typical spectral curvature of NGC 1042 ULX1, which is not revealed in any \xmm\ observations. 

The data reduction and analysis procedure are discussed in section \ref{sec:Data}. The timing and spectral analysis results of ULX1 are described in section \ref{sec:Results}. Section \ref{sec:Discussions} contains a comprehensive discussion of the results.

\begin{table*}
\centering
\caption{\label{tab:table1} Table for observation log of NGC 1042 ULX1. The cleaned exposure time from spectral data are mentioned for MOS1/MOS2/pn for \xmm\ and FPMA/FPMB for \nustar\ in ksec unit}
\begin{tabular}{ccccc}
\hline
\noalign{\smallskip}
Serial No.&Obs. ID& Date & Epoch ID & Exposure \\
\noalign{\smallskip}
\hline
\noalign{\smallskip}
\xmm\ \\
\noalign{\smallskip}
\hline\hline
		& & & &  \\
          1 &  0093630101  &2001-08-15 & XM1 & 13.6/--/9.8\\
           2& 0306230101  & 2006-01-12 & XM2& 49.4/47.9/--\\
           3&  0553300301 &2009-01-14 & XM3 & 46.8/44.5/--\\
           4& 0553300401 &2009-08-12 & XM4 & 53/--/42.6\\
           5& 0790980101 & 2017-01-17 & XM5&48.5/46/-- \\
           6&  0865260301& 2021-01-29 & XM6&--/19.9/16.3 \\
           7& 0891800401  &2021-07-16 & XM7& 26.4/26.4/20.5  \\

\hline
\noalign{\smallskip}
\nustar\ \\
\noalign{\smallskip}           
\hline\hline
& & & &    \\
   1  &  30701004002   & 2021-12-16  & N1 &104/103 \\

\hline
\end{tabular}
\end{table*}

\section{Data Reduction and analysis} \label{sec:Data}
We study NGC 1042 ULX1 by utilizing archival \xmm\ \citep{XMM} data and new \nustar\ \citep{NuSTAR} observation. The observation log is given in table \ref{tab:table1}. The archival \xmm\ observations targeted other sources like NGC 1052 galaxy and SDSSJ024052-082827. Hence, in most of the observations, ULX1 is highly off-axis. Thus, it is not always simultaneously detected by all three EPIC cameras. The new \nustar\ observation was a part of \nicer+\nustar\ joint venture. However, the \nicer\ spectra are dominated by background. Hence, we utilize \nustar\ data for scientific purposes.

The \xmm\ data are extracted using standard \xmm\ SAS software v20.0.0 \footnote{\url{https://www.cosmos.esa.int/web/xmm-newton/sas-threads}}. The EPIC data are reprocessed with \texttt{epproc} and \texttt{emproc} tasks for pn and MOS respectively. The data are cleaned from background and proton flare by eliminating the time intervals strongly affected by flaring activities by identifying those intervals in the single event high energy light curves \footnote{\url{https://www.cosmos.esa.int/web/xmm-newton/sas-thread-epic-filterbackground}}.  Thus while cleaning the data for all observations, we create good time intervals for each camera.  After background filtering, the cleaned events are utilized to extract spectra and light curves with the \texttt{evselect} tool.  The filtering expressions include \texttt{PATTERN$<=4$} for pn, which takes single and double events, and \texttt{PATTERN$<=12$} for both MOS, which take single, double, triple, and quadruple events.  For spectral analysis, we use a strict filtering constraint of \texttt{FLAG$==0$} for both pn and MOS.  For timing analysis, we perform barycentric corrections on the events using the \texttt{barycen} tool.  The background-corrected source light curves are created by \texttt{epiclccorr} task, which considers various effects like vignetting, quantum efficiency, bad pixels, and PSF variations.  For most of the \xmm\ observations, as stated earlier, the source ULX1 (RA:02 40 25.6 DEC:-08 24 30.0 ; \citealt{Sutton2012}) is highly off-axis. However, a $30$ arcsec radius circle covers the whole source in all cases.  Hence, we treat ULX1 as a point source in all observations.  We extract source photons from a $30$ arcsec circle around the source, and corresponding background photons are selected from a nearby source-free $60$ arcsec circle in the same chip.  The spectra are grouped using the task \texttt{specgroup} to have a minimum of $20$ counts per energy bin and ensure that the minimum width of a group is $1/3$ of the corresponding energy resolution (in full-width half maxima).  RMFs and ARFs are created using {\texttt{rmfgen}} and {\texttt{arfgen}} tasks respectively.

The \nustar\ data are reprocessed and science events are extracted using \texttt{HEASOFT} (v6.29) \footnote{\url{https://heasarc.gsfc.nasa.gov/docs/software/heasoft/}} routine \texttt{nupipeline} and \texttt{nuproducts}. The background subtraction of light curves for both FPMA and FPMB modules is done by the \texttt{lcmath} tool, and the spectra are grouped to have a minimum of $20$ counts per energy bin. The source region is selected with a $30$ arcsec circle around the source and the background of a $60$ arcsec radius circle from a nearby source-free region.

\section{Results} \label{sec:Results}

\subsection{Spectral Analysis} \label{sec:spectral_analysis}
 
 We utilize XSPEC v12.12.0 \citep{XSPEC} for spectral analysis of the ULX1 data. The neutral absorption component is modelled by \texttt{TBABS} with updated abundances \citep{Wilms} and cross-sections \citep{Verner}. The uncertainties on the measured spectral parameter are quoted with a 90\% confidence interval unless mentioned otherwise. We have utilized \texttt{cflux} model to estimate fluxes throughout the paper.
 
\begin{figure}
	\includegraphics[width=\columnwidth]{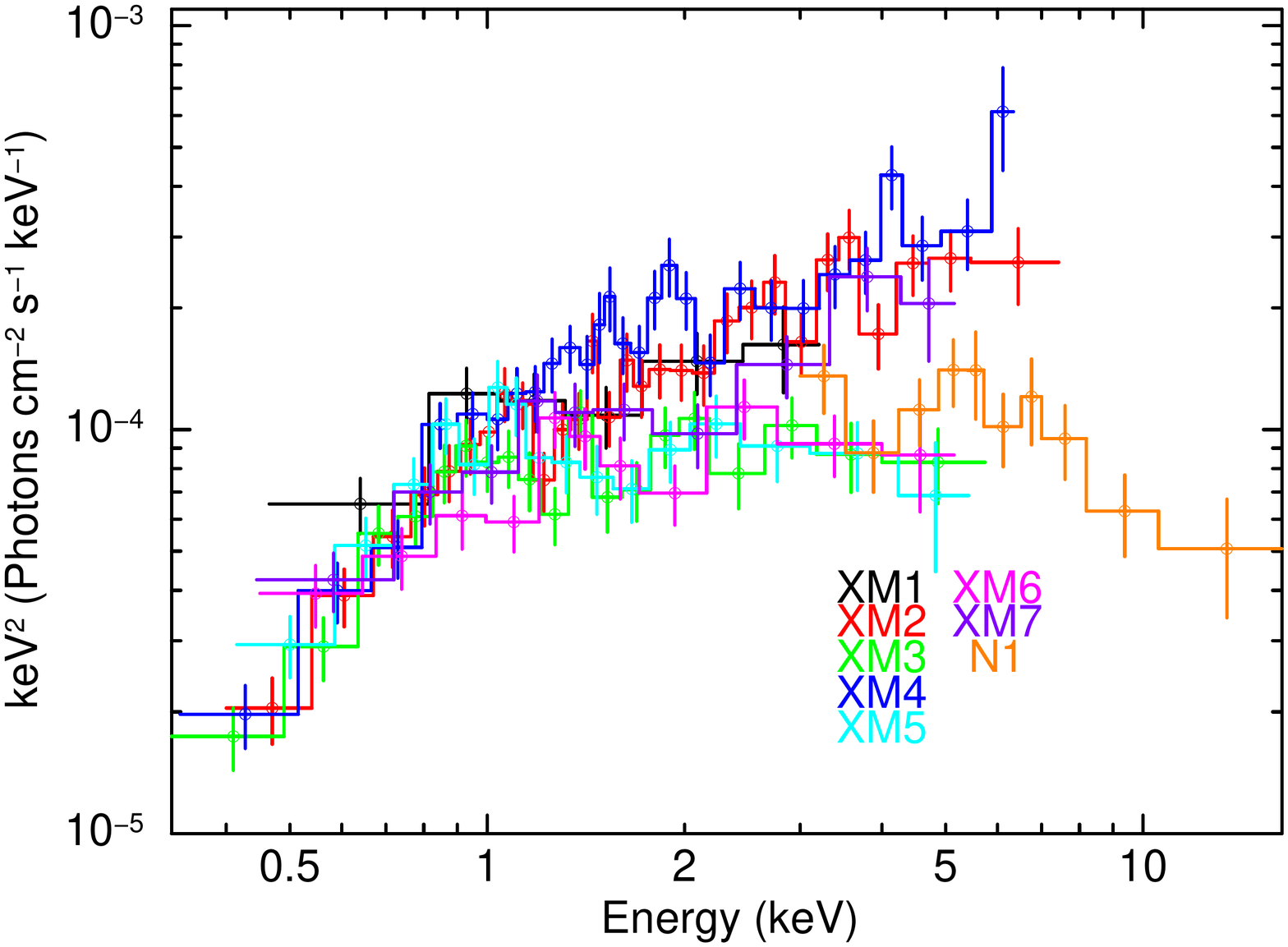}
    \caption{Unfolded ULX1 spectra for \xmm\ and \nustar\ observations. The spectra are unfolded using a zero photon index powerlaw model and have been rebinned for visual purposes. For all but XM6 epochs of \xmm\ observations, MOS1 spectra are shown here. MOS2 spectrum for XM6 and the spectrum from the FPMA module for N1 observation are presented.}
    \label{fig:eeufspec}
\end{figure}

 First, we plot the unfolded spectra from the observations we utilize in this paper to inspect the source spectral properties visually (see figure \ref{fig:eeufspec}). The unfolded spectra are generated using a powerlaw model with zero photon index, i.e., essentially a constant model. An apparent variability above $\sim 1$ keV in different epoch spectra is clear. On the other hand, below $\sim 1$ keV, the spectra remain mostly overlapping. Signal to noise ratio (S/N) decreases significantly after $8.0$ keV for \xmm\ spectra and $20.0$ keV for \nustar\ spectra. Hence analysis of \xmm\ spectra are restricted to $0.3-8.0$ keV energy range and for \nustar\ spectra $3.0-20.0$ keV energy range is utilized.
 
\begin{figure*}
	\includegraphics[width=0.66\columnwidth]{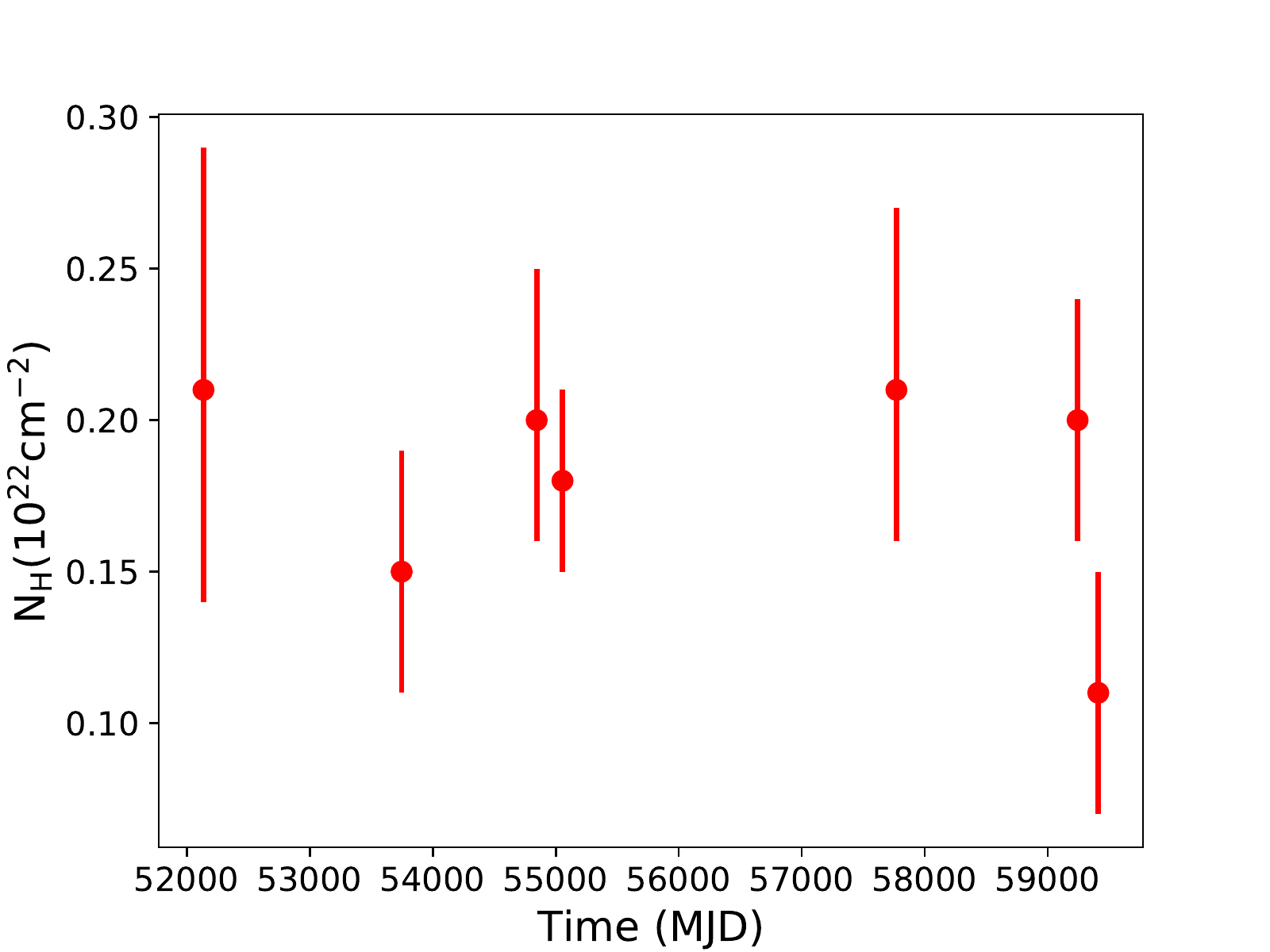}
	\includegraphics[width=0.66\columnwidth]{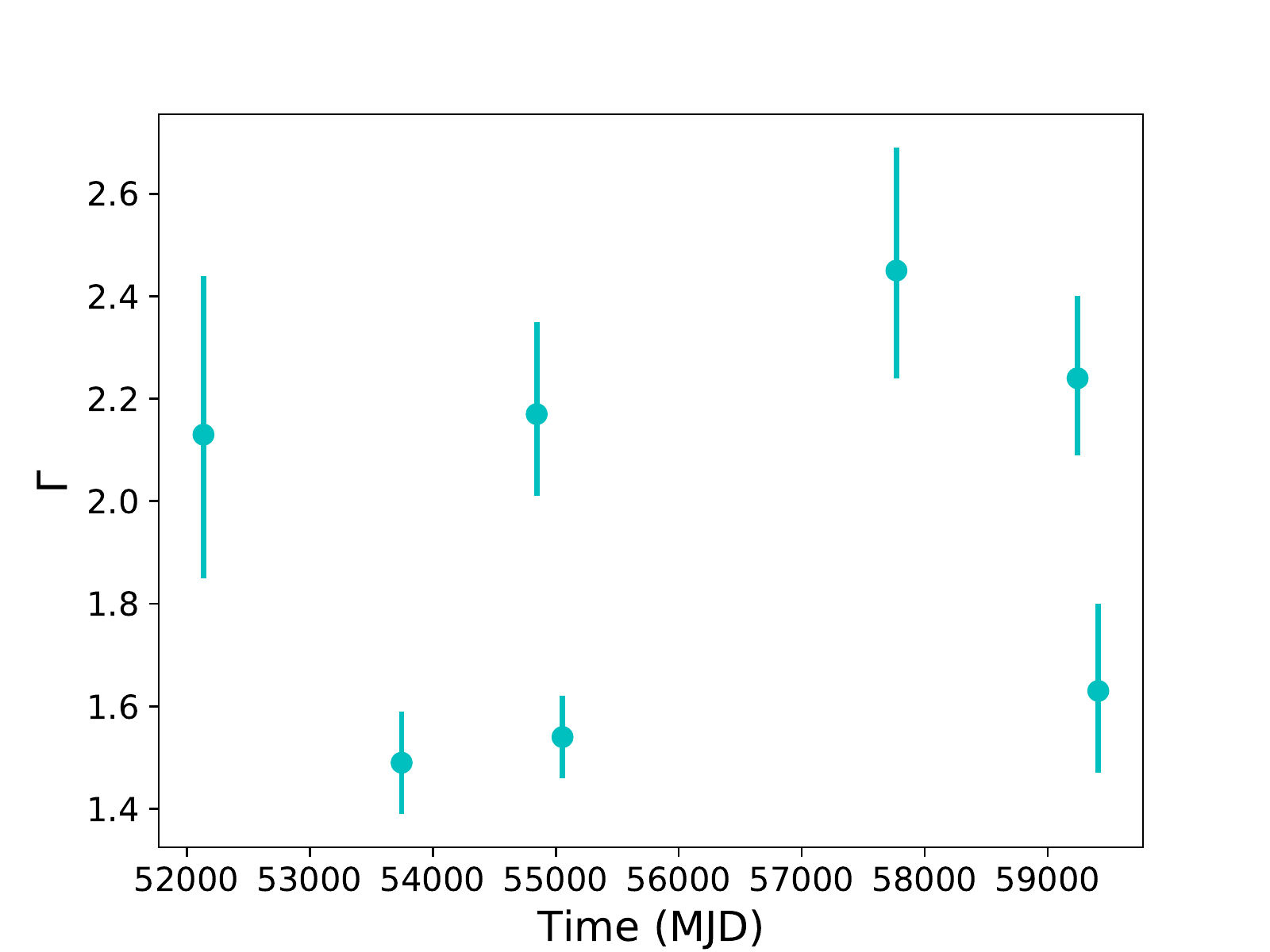}
	\includegraphics[width=0.66\columnwidth]{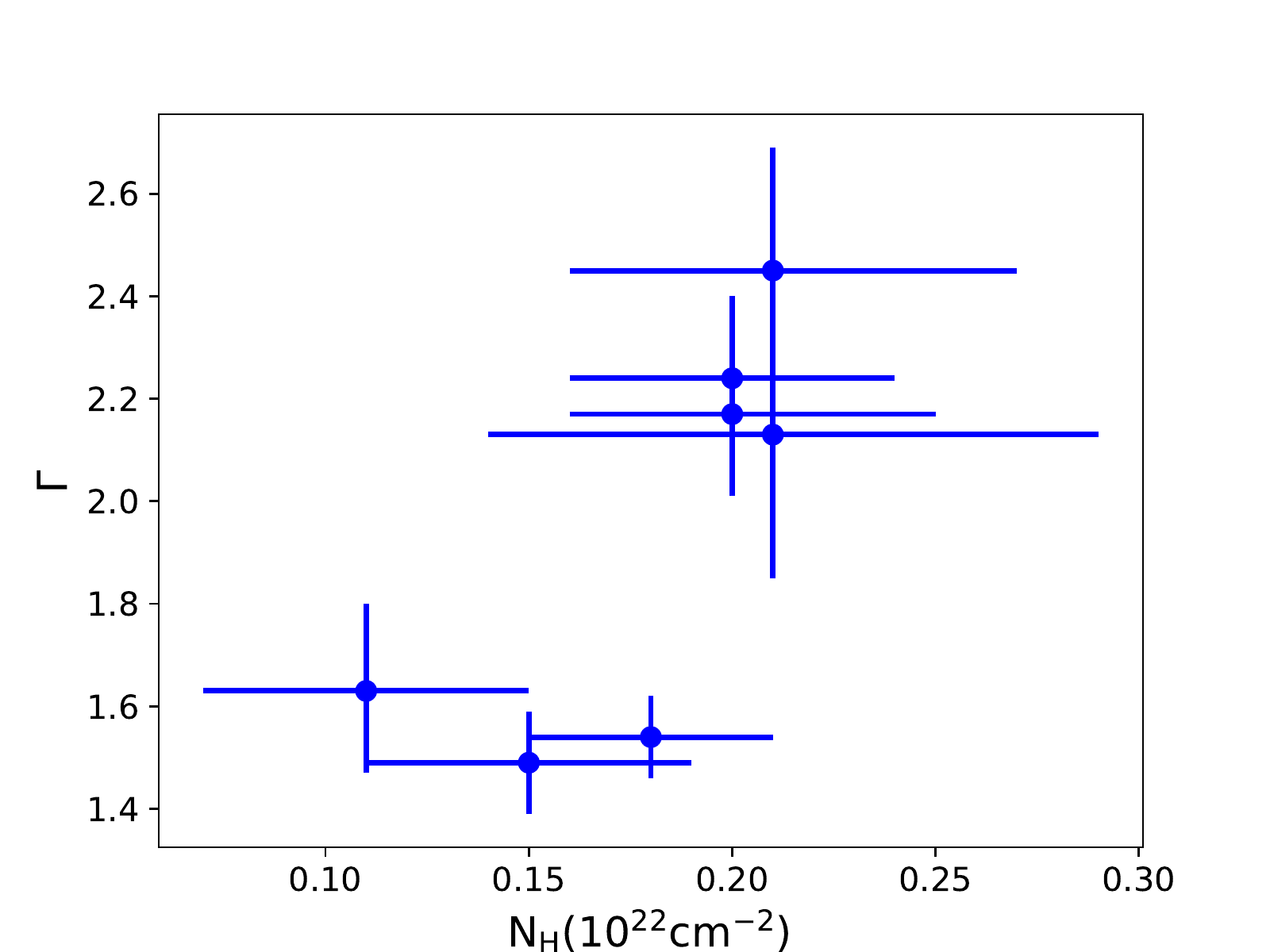}\\
    \caption{The variation of $\rm N_H$ and $\Gamma$ over different \xmm\ observation epochs are shown in left and middle panels. The relation between $\rm N_H$ and $\Gamma$ are shown in right panel. For these parameter estimates, we utilize an absorbed powerlaw model with $\rm N_H$ parameter being allowed to vary for different epochs of observation.}
    \label{fig:free_nh_parameters}
\end{figure*}

 We start analyzing data from individual \xmm\ epochs by simultaneously fitting them with a simple absorbed \texttt{powerlaw} model and observing how spectral parameters change during these epochs. Parameters from all cameras for each observation are linked except for a constant parameter that is allowed to vary to consider the cross-calibration effects. 
 Nevertheless, the parameters for each epoch are free to vary and are not linked to other epochs. The evolution of the parameters $\rm N_H$ and $\Gamma$ and their correlation are plotted in figure \ref{fig:free_nh_parameters}. We find that the $\rm N_H$ component remains mostly consistent within error among these different epochs of observation (see figure \ref{fig:free_nh_parameters} - left). However, the source shows a significant variation in spectral photon index ($\Gamma$) which ranges between $\sim 1.4-2.7$ (see figure \ref{fig:free_nh_parameters} - middle). The $\rm N_H-\Gamma$ relation is shown in figure \ref{fig:free_nh_parameters} - right panel. It shows that $\rm N_H$ does not play a significant role in the large variation of photon indices in different epochs.
 
 To see how the $\rm N_H$ influences the spectral hardness and flux, we also simultaneously fit all \xmm\ epochs spectra by keeping $\rm N_H$ free to vary globally but linking between different epochs. Keeping $\rm N_H$ linked does not significantly change the fit in terms of $\rm \chi^2/d.o.f$ from $631/534$ to $642/540$, which is expected because $\rm N_H$ in all epochs are statistically similar. Nevertheless, more importantly, this comes with a prize of better constraints on the other spectral parameters. Our goal is to quantify the spectral variability for different epochs by measuring the continuum spectral parameters. Hence, we keep this $\rm N_H$ linked for further analysis of \xmm\ data. Linking $\rm N_H$ ensures that the apparent variability is not artificial (see section \ref{sec:Discussions} for details).
 
 As stated earlier, we invoke the simplest model widely used to fit X-ray binaries and ULX spectra, an absorbed \texttt{powerlaw}. While keeping the $\rm N_H$ linked between different epochs, the best fit value of $\rm N_H$ is $0.18 \pm 0.02 \times 10^{22}$ cm$^{-2}$. The wide range variation of $\Gamma$ still holds in this case, too (see table \ref{tab:tablepow}). Both absorbed and unabsorbed luminosities are measured, which show variation in different epochs. The highest absorbed luminosity is $\sim 3$ times more than the lowest absorbed luminosity of the source.

\begin{figure*}
	\includegraphics[width=0.9\columnwidth]{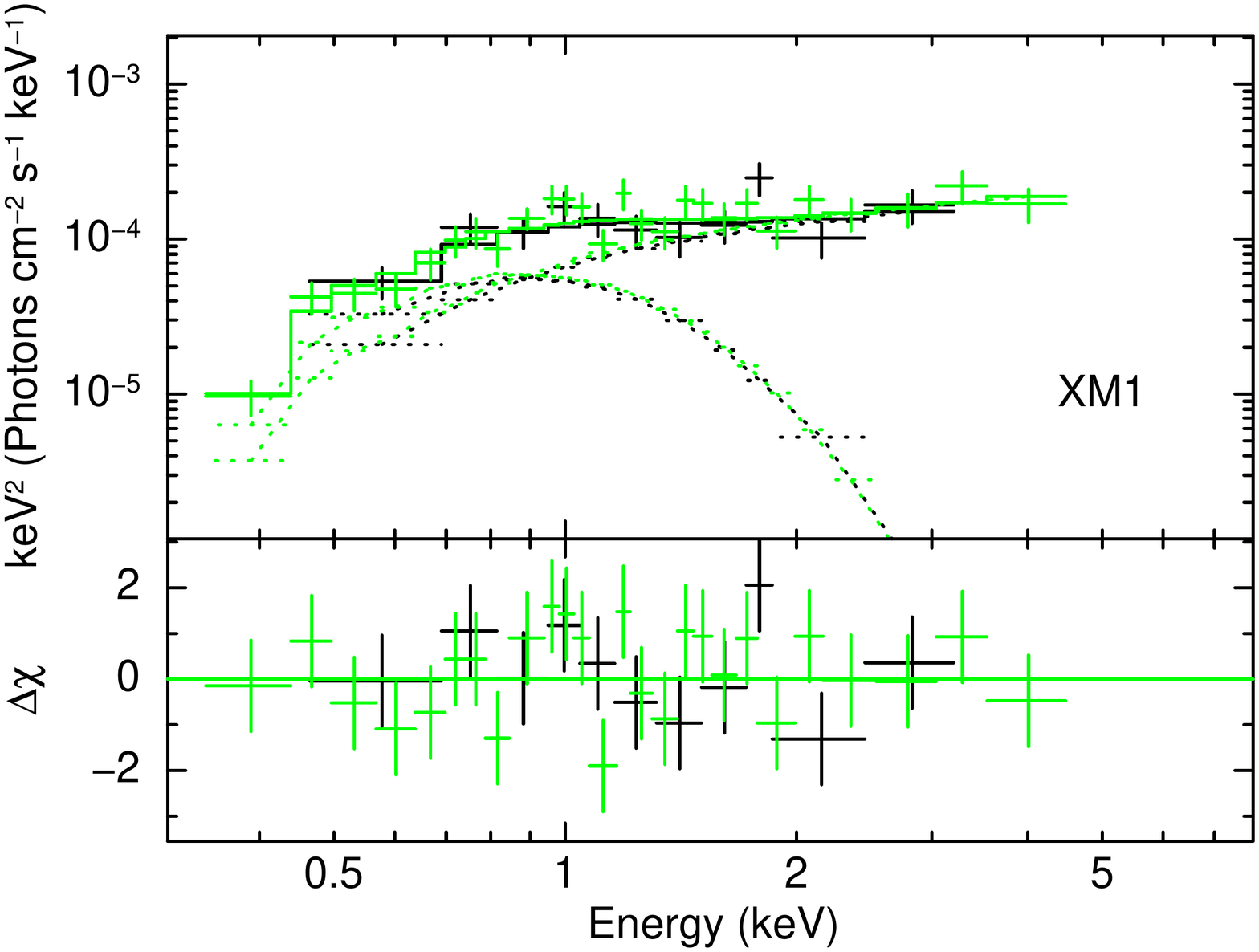}
	\includegraphics[width=0.9\columnwidth]{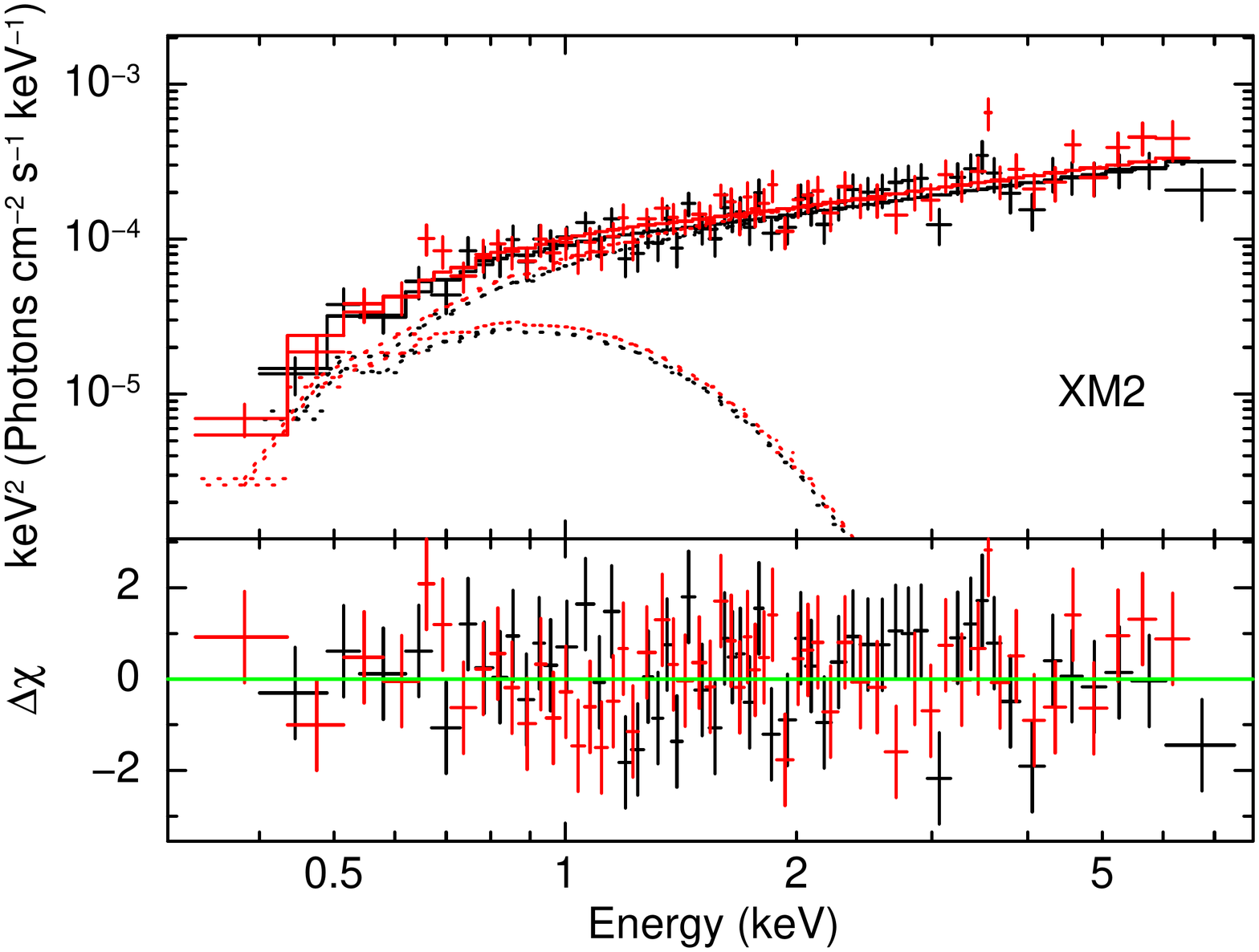}
	\includegraphics[width=0.9\columnwidth]{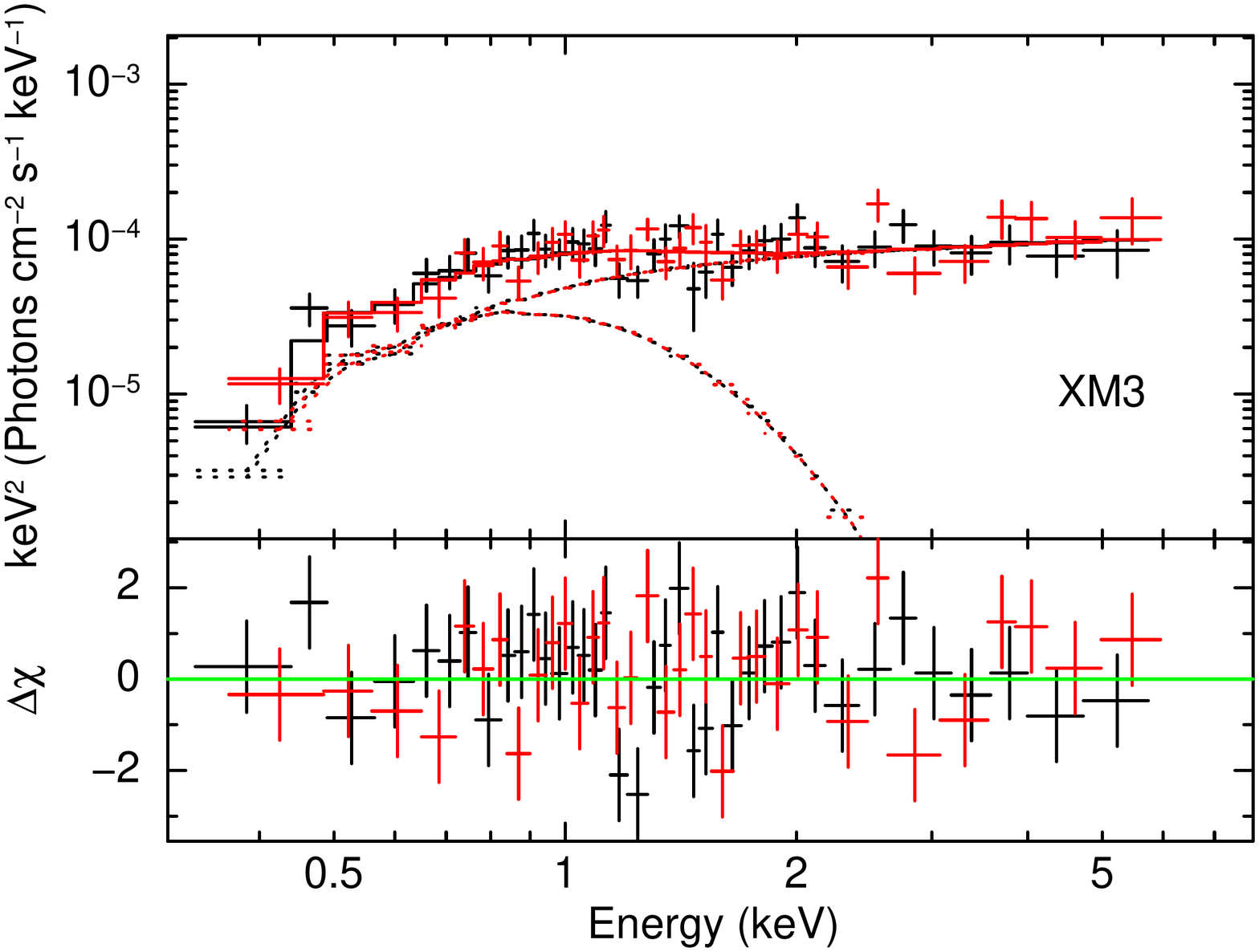}
	\includegraphics[width=0.9\columnwidth]{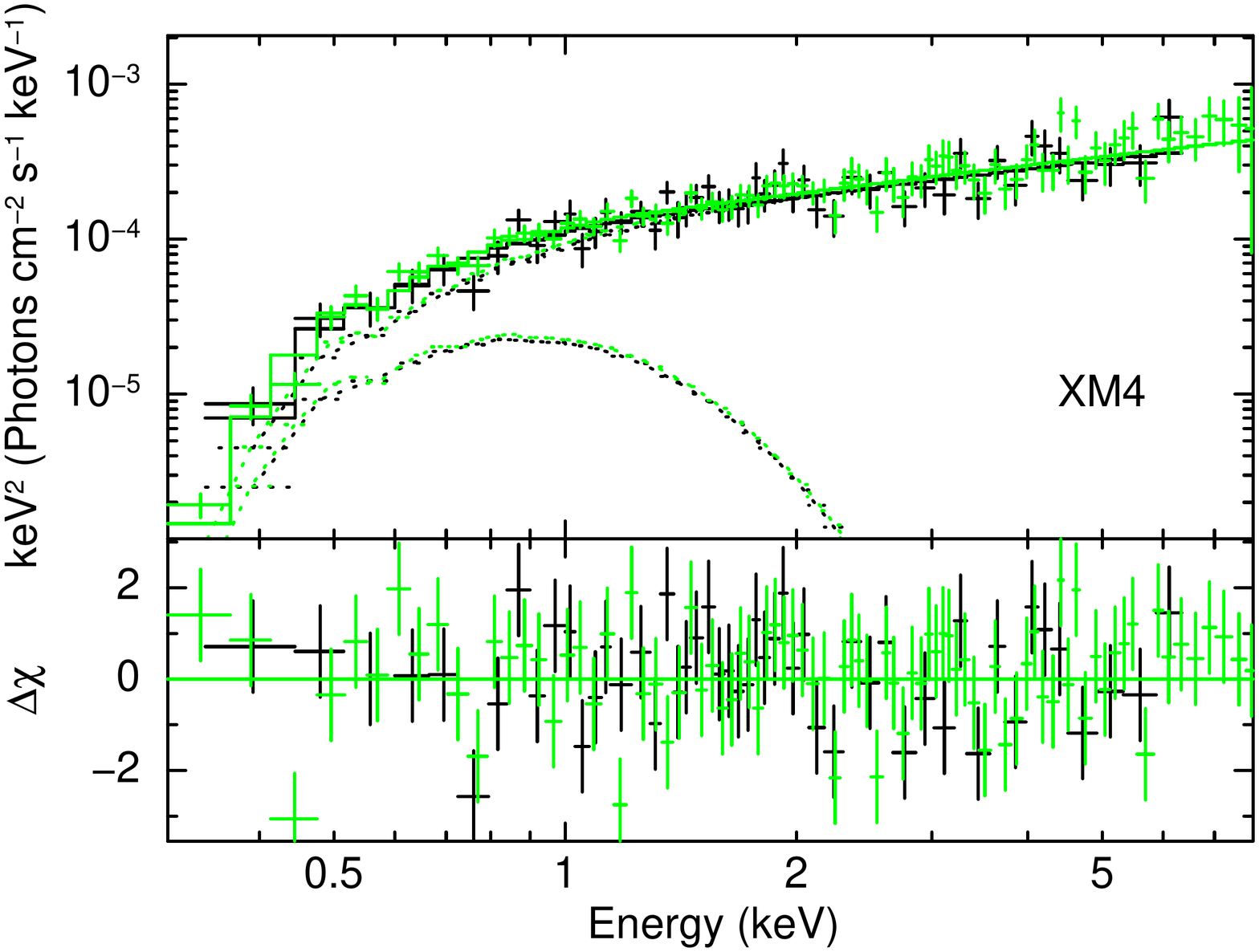}\\
	\includegraphics[width=0.9\columnwidth]{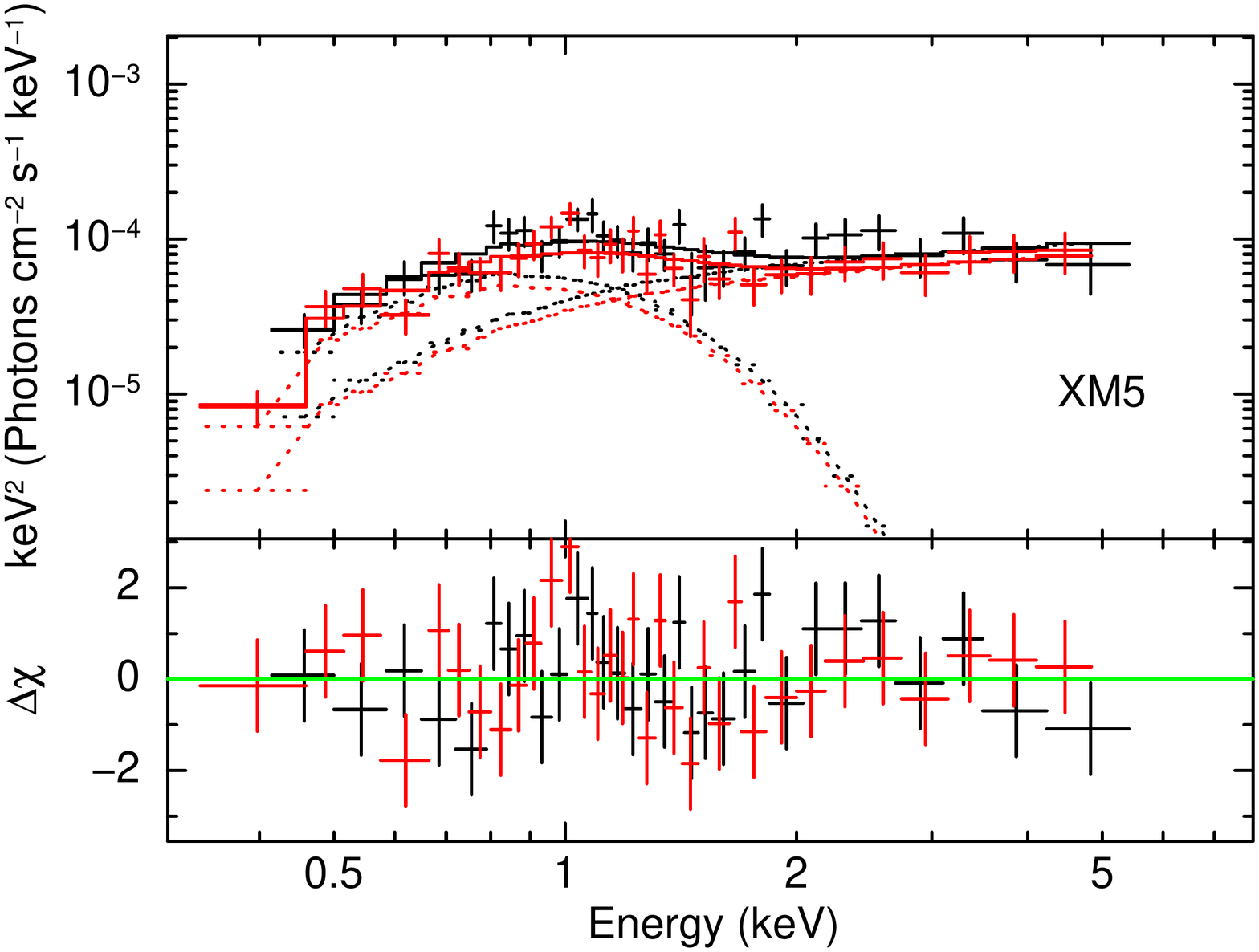}
	\includegraphics[width=0.9\columnwidth]{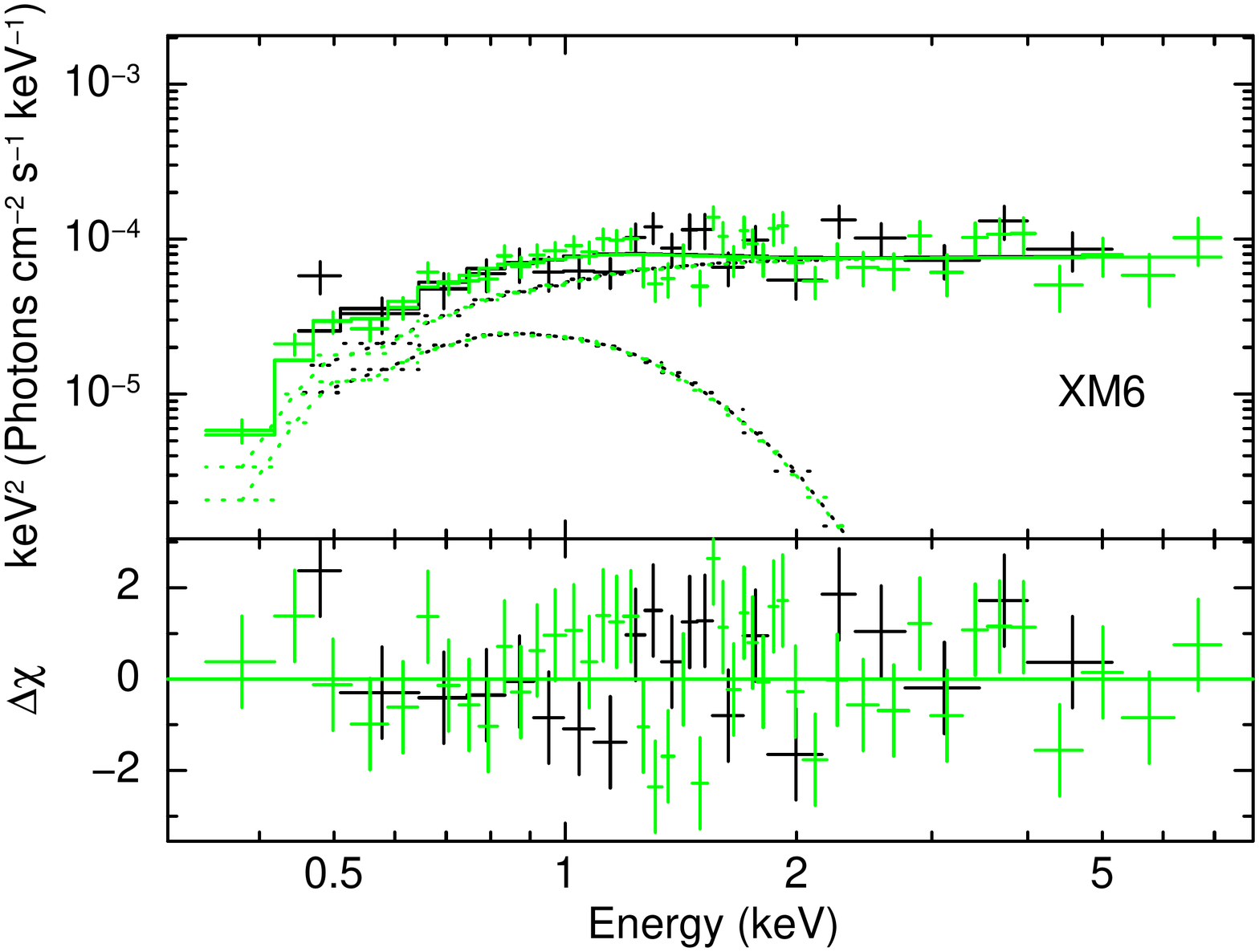}
	\includegraphics[width=0.9\columnwidth]{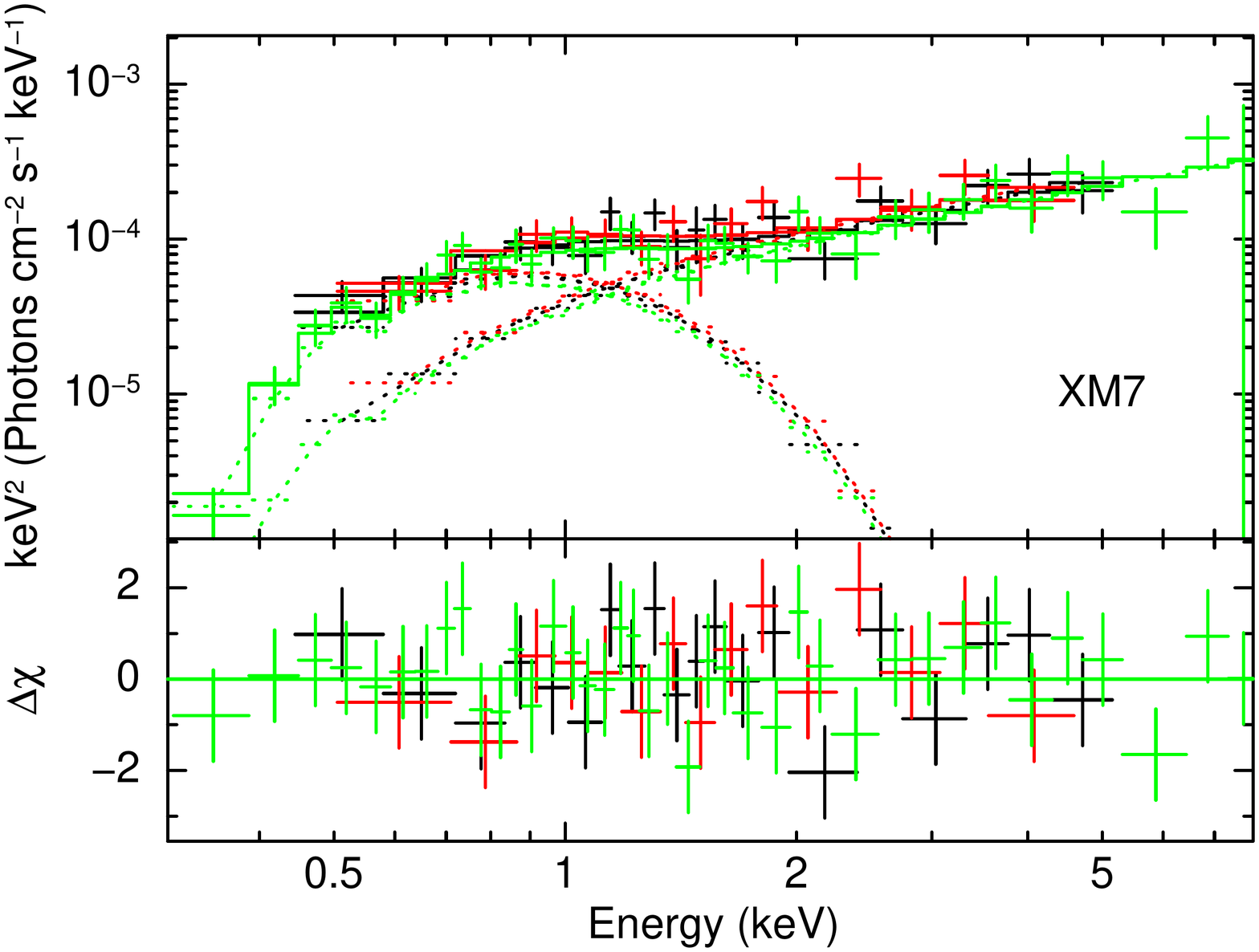}
	\includegraphics[width=0.9\columnwidth]{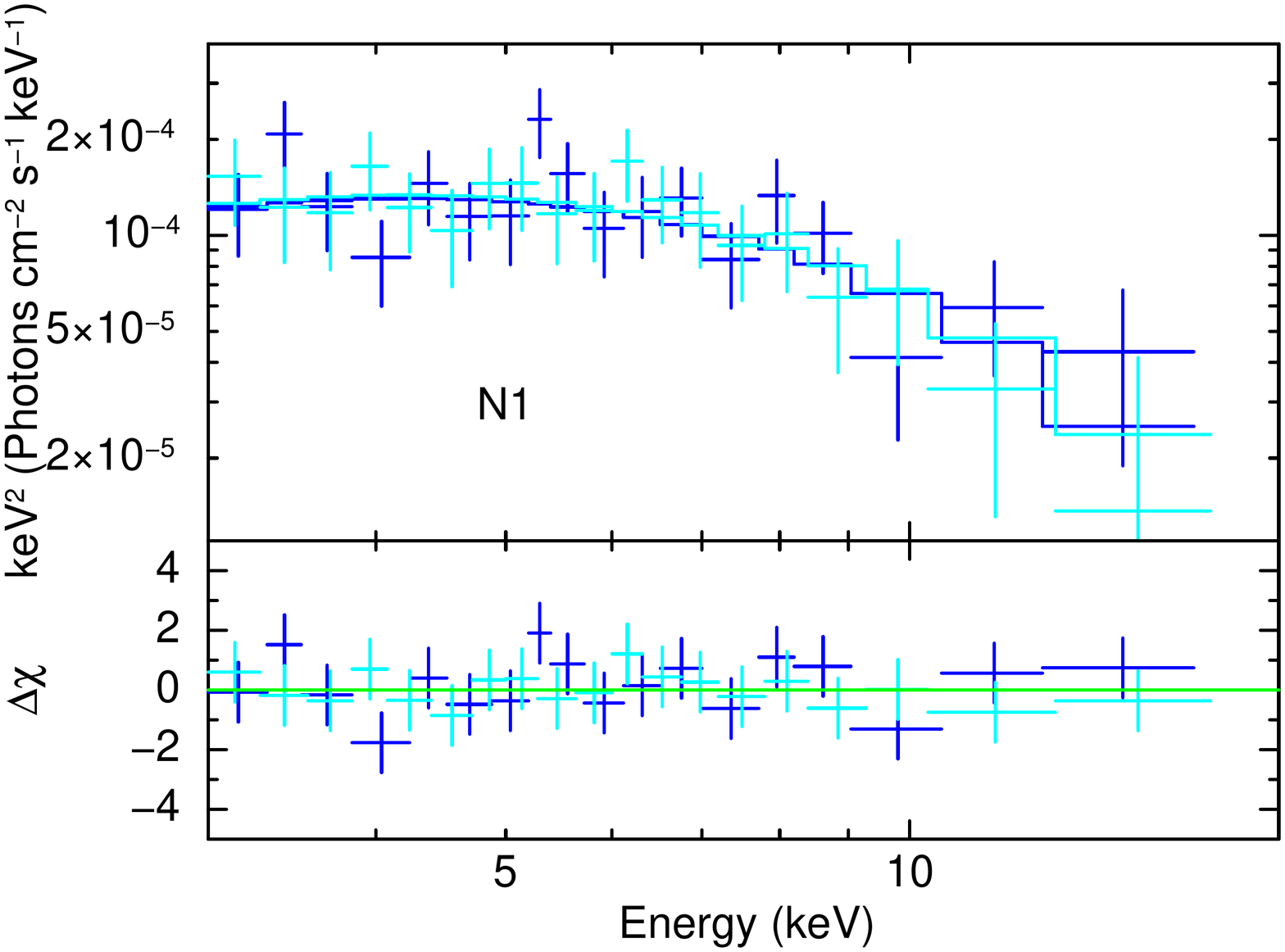}
    \caption{The spectra and residuals for \texttt{diskbb+powerlaw} model of \xmm\ observations and \texttt{cutoffpl} model of \nustar\ observation. The black, red and green colors correspond to MOS1, MOS2 and pn, while blue and light blue colors correspond to FPMA and FPMB data respectively. Corresponding additive models are also shown in top panels of each figure.}
    \label{fig:eeufspec_residuals}
\end{figure*}

 We find that a single component multi color disk model (\texttt{diskbb} in XSPEC) is not a good description for the simultaneous spectral fit ($\rm \chi^2/d.o.f = 932/540$). We then explore two-component models, often the best description of ULX spectra. We find that a \texttt{diskbb} component in addition to the \texttt{powerlaw}, improves the fit statistics significantly compared to a single component \texttt{powerlaw} fit. We also find that the temperature of the disk component remains statistically the same (within $90\%$ confidence) in all epochs; hence we link this parameter for different epochs and let it vary globally. We find that compared to the single component \texttt{powerlaw} model, the \texttt{diskbb+powerlaw} model gives much better statistical fit ($\Delta\chi^2 \simeq -68$ for $8$ less degrees of freedom). The variation in photon indices still prevails even though the cool disk component significantly contributes to the softer regime of the spectra (see the model components in figure \ref{fig:eeufspec_residuals}). We also see a trend that for XM2 and XM4 epochs, which have shown hard spectra, have comparatively more ratio of the powerlaw flux and disk flux ($\frac{F_{pl}}{F_{disk}}$) than the epochs which have softer spectra (see table~\ref{tab:tablediskpow}).
 
 We also explore another two component model widely used for ULX spectral fits, i.e. \texttt{diskbb+diskbb} (e.g., \citealt{Gurpide2021a, Koliopanos2017}) which comprises of two temperature disk blackbody components. However, we find that when we fit two \texttt{diskbb} components, in some cases the uncertainties in hotter disk component is unphysically high, this could be partly due to the degeneracy between different parameters. Hence, we do not explore this model in detail for the current data. Certainly, we should mention that future on-axis and long exposure observations of this source might be able to adequately constrain such model.
 
 With the current limitation of data, we consider that the two-component model \texttt{diskbb+powerlaw} provides a good statistical fit of ULX1 spectra in all epochs, and we consider this combination of the model as the best fit model for NGC 1042 ULX1. The residuals and unfolded spectra for all \xmm\ epochs, along with the additive model components for this best fit model combination, are shown in figure \ref{fig:eeufspec_residuals}. The spectral parameters are noted in table \ref{tab:tablepow} and \ref{tab:tablediskpow} for \texttt{powerlaw} and \texttt{diskbb+powerlaw} models.

 Apart from analyzing the archival \xmm\ observations, we analyze the new \nustar\ data to understand the spectral nature of ULX1 in the hard energy band. Unfortunately, the simultaneous soft counterpart of \nicer\ observation of ULX1 is completely background-dominated. Hence, we cannot use the \nicer\ data for any meaningful scientific analysis. Since the soft counterpart is unavailable, the \nustar\ spectra are fitted with absorption $\rm N_H$ fixed to the best fit values taken from \xmm\ fits. First, we fit an absorbed \texttt{powerlaw} model with $\rm N_H$ fixed to $0.18 \times 10^{22}$ cm$^{-2}$ and find that $\rm \chi^2/d.o.f = 32/34$ with $\Gamma = 2.74^{+0.21}_{-0.20}$, which is a statistically acceptable fit suggesting that the current \nustar\ data are broadly consistent with a simple powerlaw model. However, if we include an exponential cutoff powerlaw model (\texttt{cutoffpl}) instead of \texttt{powerlaw}, the fit provides a lower $\chi^2$ value ($\rm \chi^2/d.o.f = 21/33$). However, due to the limited S/N of the data, the photon index has a large error bar ($\Gamma = 0.14^{+1.39}_{-1.65}$), including a negative value in the lower error and an unconstrained normalization ($<15.64 \times 10^{-5}$) with a folding energy value of $\rm E_{fold} = 2.45^{+2.86}_{-0.98}$ keV. Hence, we freeze the photon index of \texttt{cutoffpl} model to $0.59$, which is typical value for pulsar ULXs (see \citealt{Walton2020}). The fit remains statistically similar to a free photon index. The $\rm \chi^2/d.o.f = 21/34$ with folding energy at $\rm E_{fold} = 2.96^{+0.39}_{-0.33}$ keV. This folding energy value has to be treated with caution when compared with other ULXs. Due to the unavailability of simultaneous soft counterpart data, the \texttt{cutoffpl} parameters, including the photon index, are not well determined. Nevertheless, the turnover in the \nustar\ spectra is apparent from figure \ref{fig:eeufspec}. The archival \xmm\ data could not detect the cutoff in any observation when fitted with \texttt{cutoffpl} model (i.e., unconstrained folding energy with no statistical improvement compared to simple \texttt{powerlaw} fit), whereas, the new \nustar\ data detect the cutoff, although at the low energy threshold of \nustar. The unabsorbed flux in $3.0-20.0$ keV energy range is $(2.49^{+0.30}_{-0.29}) \times 10^{-13}$ \fluxcgs and corresponding luminosity is $(1.06^{+0.13}_{-0.12}) \times 10^{+40}$ \lumcgs. The \nustar\ spectra are plotted in figure \ref{fig:eeufspec_residuals} (last panel). In order to provide statistical justification for the observed spectral turnover, we further fit the \nustar\ data with a broken powerlaw model following the work by \citet{Stobbart2006}. We find that the $\rm \chi^2/d.o.f = 19/32$ and the break energy is $\rm E_{break} = 6.64^{+1.98}_{-1.24}$ keV, with power law photon index for $\rm E<E_{break}$ is $1.98^{+0.49}_{-0.60}$ and for $\rm E>E_{break}$ is $4.0^{+2.49}_{-0.77}$. This indeed statistically validate the observed spectral break in the \nustar\ data.

\begin{table*} 
\centering 
 \caption{Parameter table for \texttt{powerlaw} model in seven epochs of \xmm\ observation of NGC 1042 ULX1 for linked $N_H$. We list the observed (absorbed) fluxes and luminosities in 0.3-8.0 keV energy range. The intrinsic (unabsorbed) fluxes are typically $\sim 1.2$--$1.6$ times of the observed flux.}
%\begin{adjustbox}{width=1.0\textwidth} 
 \begin{tabular}{lcccccccccr} 
\hline 
Parameters & Unit  &  XM1& XM2 & XM3 &  XM4 & XM5 & XM6 & XM7  \\
 \hline
 
\hline 
 \multicolumn{9}{c|}{Model = \texttt{TBabs*powerlaw}}\\
 \hline 
 \hline 
$\rm N_H$ & $10^{22} cm^{-2}$ & $0.18 \pm 0.02$ &   &   &    &   &  &  \\
$\Gamma$ &  & $2.00 \pm 0.14$ & $1.55 \pm 0.07$ & $2.11 \pm 0.10$ &  $1.53 \pm 0.06$ & $2.34^{+0.12}_{-0.11}$ & $2.17 \pm 0.09$ & $1.84^{+0.12}_{-0.11}$ \\
 $\rm N_{pl}$ & $10^{-4} $ & $1.51^{+0.20}_{-0.19}$ & $1.19 \pm 0.08$ & $1.01 \pm 0.08$ &  $1.46 \pm 0.10$ & $1.18 \pm 0.10$ & $0.98 \pm 0.10$ & $1.20 \pm 0.13$  \\
 \hline
 $\rm \chi^2/d.o.f$  & & $642/540$  &  &   &   &  &  &   \\
 \hline
 $\rm F_{obs}$ & $10^{-13} \fluxcgs$ & $5.72 ^{+ 0.81 }_{- 0.76 }$ & $6.96 ^{+ 0.44 }_{- 0.43 }$ & $3.54 ^{+ 0.28 }_{- 0.27 }$ &  $8.77 ^{+ 0.53 }_{- 0.52 }$ & $3.55 ^{+ 0.31 }_{- 0.30 }$ & $3.28 ^{+ 0.32 }_{- 0.31 }$ & $5.26 ^{+ 0.59 }_{- 0.57 }$\\
 $\rm L_{obs}$ & $10^{+40} \lumcgs$ & $2.44 ^{+ 0.35 }_{- 0.32 }$ & $2.98 ^{+ 0.19 }_{- 0.18 }$ & $1.51 ^{+ 0.12 }_{- 0.11 }$ &  $3.75 \pm 0.22$ & $1.52 \pm 0.13$ & $1.40 ^{+ 0.14 }_{- 0.13 }$ & $2.25 ^{+ 0.25 }_{- 0.24 }$ \\
 \hline
 
 \label{tab:tablepow}
 \end{tabular} 
%\end{adjustbox} 
\end{table*}

\begin{table*} 
\centering 
 \caption{Parameter table for \texttt{diskbb+powerlaw} model in seven epochs of \xmm\ observation of NGC 1042 ULX1. We list the observed (absorbed) fluxes and luminosities in 0.3-8.0 keV energy range. We also note the individual intrinsic flux of additive components in the same energy range.}
%\begin{adjustbox}{width=1.0\textwidth} 
 \begin{tabular}{lcccccccccr} 
\hline 
Parameters & Unit  &  XM1 & XM2 & XM3 &  XM4 & XM5 & XM6 & XM7  \\
 \hline
 
\hline 
\multicolumn{9}{c|}{Model = \texttt{TBabs(diskbb+powerlaw)}}\\
 \hline 
 \hline 
$\rm N_H$ & $10^{22} cm^{-2}$ & $0.23^{+0.04}_{-0.03}$ &   &  &    &   &  &   \\
$\rm T_{in}$ & keV & $0.23^{+0.04}_{-0.03}$ &   &   &    &   &  &   \\
$\rm N_{disk}$ &  & $6.20^{+10.31}_{-4.45}$ & $2.82^{+5.78}_{-1.94}$  & $3.67^{+6.42}_{-2.41}$  & $2.45^{+5.60}_{-1.82}$   & $6.46^{+9.50}_{-3.48}$  & $2.70^{+5.66}_{-2.30}$ & $6.38^{+9.91}_{-3.47}$  \\
$\Gamma$ &  & $1.56^{+0.49}_{-0.47}$ & $1.39 \pm 0.13$ & $1.81 \pm 0.28$ &  $1.45 \pm 0.11$ & $1.71^{+0.36}_{-0.39}$ & $2.01^{+0.27}_{-0.26}$ & $1.10^{+0.24}_{-0.25}$ \\
 $\rm N_{pl}$ & $10^{-4} $ & $0.99^{+0.51}_{-0.37}$ & $0.99^{+0.15}_{-0.14}$ & $0.72^{+0.23}_{-0.20}$ &  $1.33 \pm 0.16$ & $0.60^{+0.27}_{-0.22}$ & $0.79^{+0.26}_{-0.20}$ & $0.58^{+0.18}_{-0.15}$  \\
 \hline
 $\rm \chi^2/d.o.f$  &  & $574/532$  &    &   &   &  &  &   \\
 \hline
 Spectral regimes &  & HUL/SUL & HUL  & HUL/SUL & HUL  & HUL/SUL  & HUL/SUL & HUL  \\
 \hline
 $\rm F_{obs}$ & $10^{-13} \fluxcgs$ & $6.50 ^{+ 1.63 }_{- 1.27 }$ & $7.24 ^{+ 0.49 }_{- 0.48 }$ & $3.69 ^{+ 0.36 }_{- 0.34 }$ &  $8.88 ^{+ 0.57 }_{- 0.56 }$ & $3.85 ^{+ 0.44 }_{- 0.40 }$ & $3.25 ^{+ 0.36 }_{- 0.34 }$ & $6.67 ^{+ 0.89 }_{- 0.82 }$\\
 $\rm L_{obs}$ & $10^{+40} \lumcgs$ & $2.78 ^{+ 0.70 }_{- 0.54 }$ & $3.10 \pm 0.21$ & $1.58 ^{+ 0.15 }_{- 0.14 }$ &  $3.80 ^{+ 0.25 }_{- 0.24 }$ & $1.65 ^{+ 0.19 }_{- 0.17 }$ & $1.39 \pm 0.15$ & $2.85 ^{+ 0.38 }_{- 0.35 }$ \\
 $\rm F_{pl}$ & $10^{-13} \fluxcgs$ & $6.87 ^{+ 1.08 }_{- 1.04 }$ & $8.04 ^{+ 0.51 }_{- 0.5 }$ & $4.2 ^{+ 0.7 }_{- 0.58 }$ &  $10.18 ^{+ 0.6 }_{- 0.6 }$ & $3.73 ^{+ 0.8 }_{- 0.64 }$ & $4.14 ^{+ 0.93 }_{- 0.67 }$ & $6.36 ^{+ 0.76 }_{- 0.73 }$\\
 $\rm F_{disk}$ & $10^{-13} \fluxcgs$ & $2.57 ^{+ 1.36 }_{- 1.75 }$ & $1.17 ^{+ 0.68 }_{- 0.61 }$ & $1.53 ^{+ 0.74 }_{- 0.84 }$ &  $1.02 ^{+ 0.73 }_{- 0.63 }$ & $2.68 ^{+ 0.83 }_{- 0.86 }$ & $1.12 ^{+ 0.78 }_{- 0.92 }$ & $2.65 ^{+ 0.8 }_{- 0.64 }$\\
 \hline
 
 \label{tab:tablediskpow}
 \end{tabular} 
%\end{adjustbox} 
\end{table*}

\subsection{Timing Analysis}
\begin{figure}
	\includegraphics[width=\columnwidth]{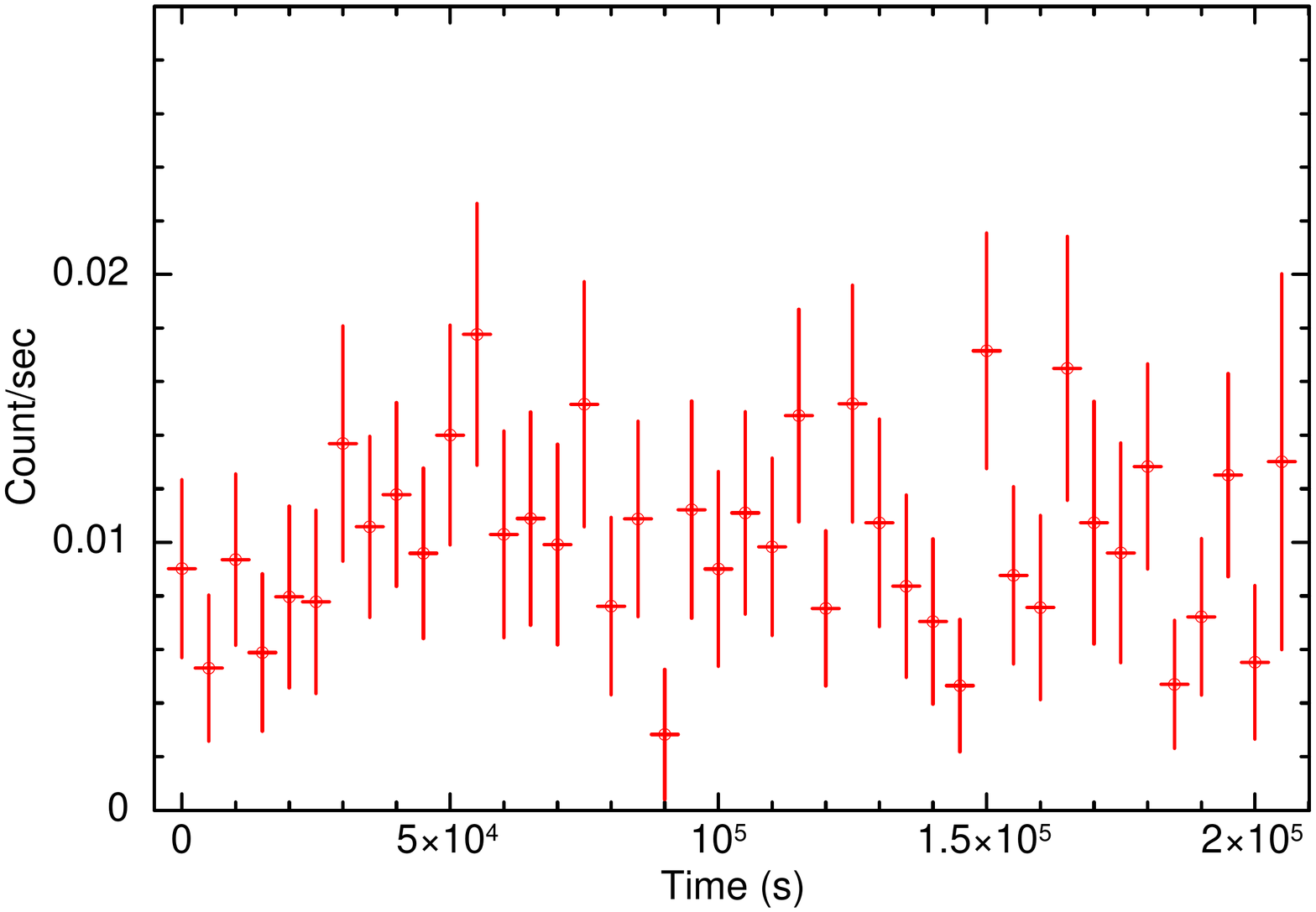}
    \caption{Representative $3.0-20.0$ keV background subtracted cleaned light curve from \nustar\ FPMA with a binning of 5000 seconds.}
    \label{fig:lcurve}
\end{figure}

NGC 1042 ULX1 shows steady nature in the time series of all \xmm\ and \nustar\ observations. The background-subtracted source light curve for FPMA is plotted in figure \ref{fig:lcurve}. We utilize the pn data of four \xmm\ observations to search for pulsation using \texttt{HENDRICS v7.0.2} software \citep{HENDRICS, Stingray} implementing the \texttt{HENzsearch} tool with the fast folding algorithm, which searches for the first spin derivative. We have restricted the search to the frequency range of $0.1-6.8$ Hz and the energy range of $0.3-8.0$ keV. No pulsation is detected in any observation. We obtain the upper limits of pulsed amplitude for all four epochs, which range between $\sim 20-40\%$ (in $90\%$ confidence interval). We also utilize the new \nustar\ data for timing analysis and search for pulsation with a similar method within the $0.1-10.0$ Hz frequency range and $3.0-20.0$ keV energy range. However, the S/N of \nustar\ data is low. Hence, we do not detect any pulsation here either. The upper limit of pulsed amplitude is $\sim 40\%$.

\section{Discussions} \label{sec:Discussions}
This paper aims to study a luminous and variable ULX source, NGC 1042 ULX1. \citet{Sutton2012} studied a sample of extremely bright ULX sources; NGC 1042 ULX1 was one of them. Both possibilities of sub-Eddington accretion onto the intermediate-mass black hole and super-Eddington stellar-mass accretors scenario are discussed in that paper, which analyzed three archival \xmm\ observations and one \chandra\ observation of ULX1. However, the absence of apparent soft excess and characteristic spectral cutoff in \xmm\ data favoured the IMBH scenario. The spectral curvature of NGC 1042 ULX1, which is not constrained by any of the \xmm\ observations, can be due to the low quality of the data (see the discussion of \citealt{Sutton2012}). Nevertheless, the current analysis of \xmm\ and \nustar\ data provides more insight into the source, thoroughly discussed in this section.

\subsection{Accretion states of ULX1} \label{sec:accretion_state}
It is important to compare the accretion state of any individual ULX with the known ULX spectral states (e.g., \citealt{Sutton2013,  Kaaret2017, Gurpide2021a, Gurpide2021b}). Based on the definition followed by \citealt{Sutton2013} for different ULX states and interpreting the relative contribution of soft and hard components for NGC 1042 ULX1 indicates similarity with the ultraluminous state. The spectral hardness mostly indicates the source resembles HUL regime, although, in some observations, the error on the photon index extends to the SUL regime. Due to the limited quality of \xmm\ data, it would be difficult to rule out the degeneracy between the soft thermal disk component and hard powerlaw component, which is manifested by the somewhat large measurement uncertainties in the spectral parameters. Nevertheless, the changes in spectral profile are evident from figure \ref{fig:eeufspec}, figure \ref{fig:eeufspec_residuals} and the quantified results noted in table \ref{tab:tablepow} and \ref{tab:tablediskpow}.

Another critical point to note is that the spectral differences are more prominent beyond $\sim 1$ keV. Such an interesting behavior has also been reported in other ULXs, like NGC 1313 X1, NGC 55 ULX1, Holmberg IX X1, M51 ULX8, NGC 4395 ULX1 \citep{Sutton2013, Middleton2015, Walton2020, Gurpide2021a, Ghosh2022}. In general, it can be understood as a scenario where the cool emission component does not vary in different observations, but the hot counterpart exhibits variability. Typically for ULXs, it is understood that the cool disk blackbody component is the manifestation of optically thick wind which is launched near the spherization radius when the accretion rate reaches the Eddington limit and the powerlaw component approximates a hotter inner accretion flow modified by a Comptonization process, which is a dominant radiative mechanism in many ULXs (see e.g., \citealt{Urquhart2016, Pinto2017, Walton2020}). Current data do not allow us to constrain a theoretical Comptonization model like \texttt{comptt}. Since the spectral hardness for a Comptonization process directly depends on how much photons are up-scattered from the seed photons of the disk, the powerlaw model as an approximation indicates that there is a variation in the up-scattered photon fraction in different epochs of \xmm\ if the hard spectrum is indeed dominated by Comptonizaion process. It is also important to note that the temperature of the soft \texttt{diskbb} component for ULX1 remains similar within the $\sim 90\%$ confidence interval, again suggesting that the spectral variability originates from the hard component i.e., either variability in inner accretion flow or contribution from Comptonization process.

\citealt{Gurpide2021a, Gurpide2021b} studied spectral variability in a sample of ULX sources and predicted physical scenarios which could generate such variabilities in these sources. One crucial understanding is that, for super-Eddington stellar-mass accretors, a strong radiatively driven outflow is generated near the spherization radius due to the high accretion rate in the system. The outflow can be optically thin or thick depending on whether the inclination angle of the system is low or high, respectively \citep{Gurpide2021a, Poutanen2007}. Thus, a low inclination of the disk would enhance the probability of hard photons dominating the line of sight emission. On the contrary, for a higher inclination angle, the hot part of the disk would be obscured, and most of the hard photons would be down-scattered by the optically thick wind, and thus soft emission would dominate the spectrum. Thus, changes in inclination would imply the variation in occultation of the inner region of the disk which would imprint the variability in hardness of the observed spectrum.

\begin{figure*}
    \includegraphics[width=\columnwidth]{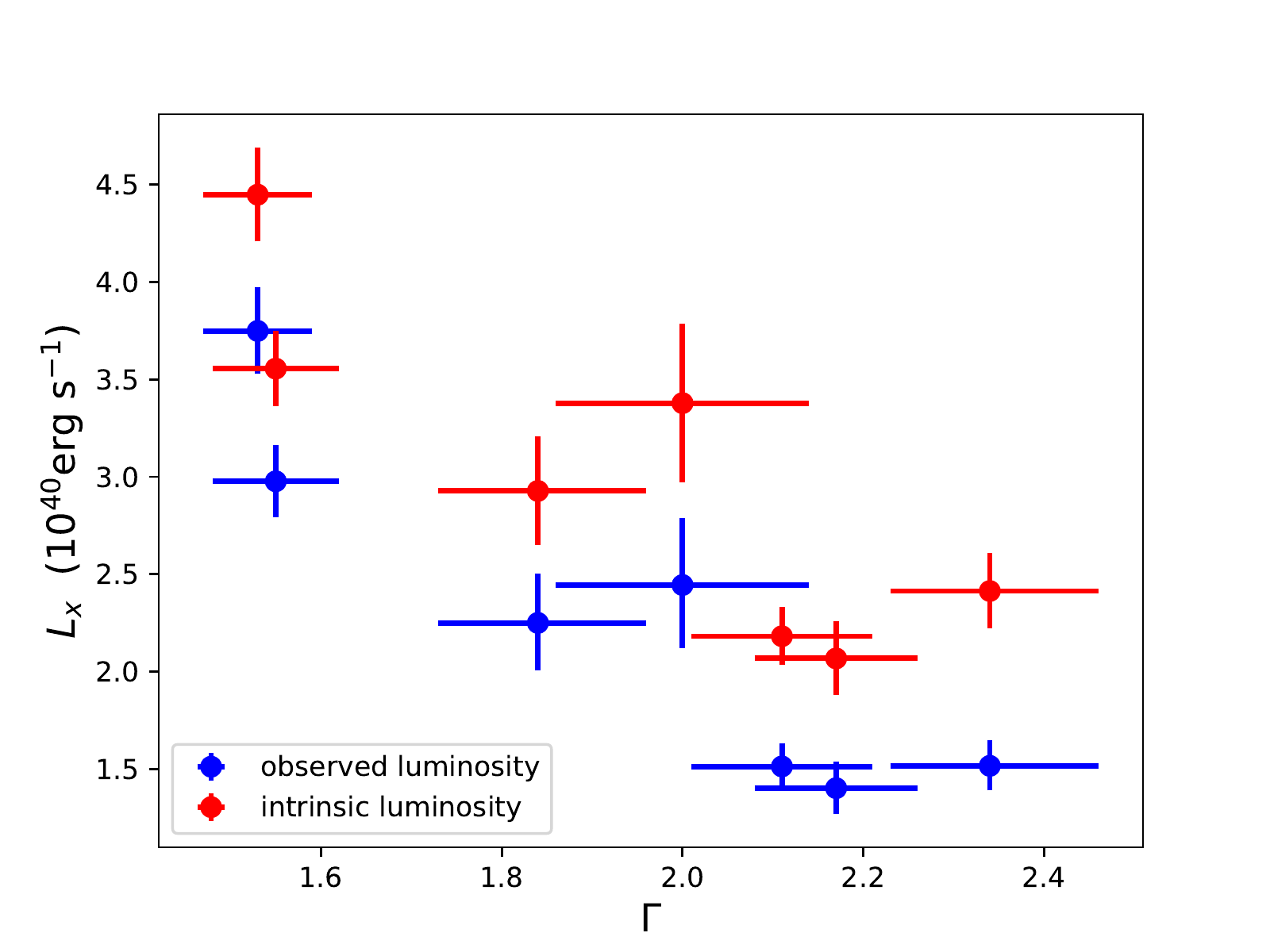}
	\includegraphics[width=\columnwidth]{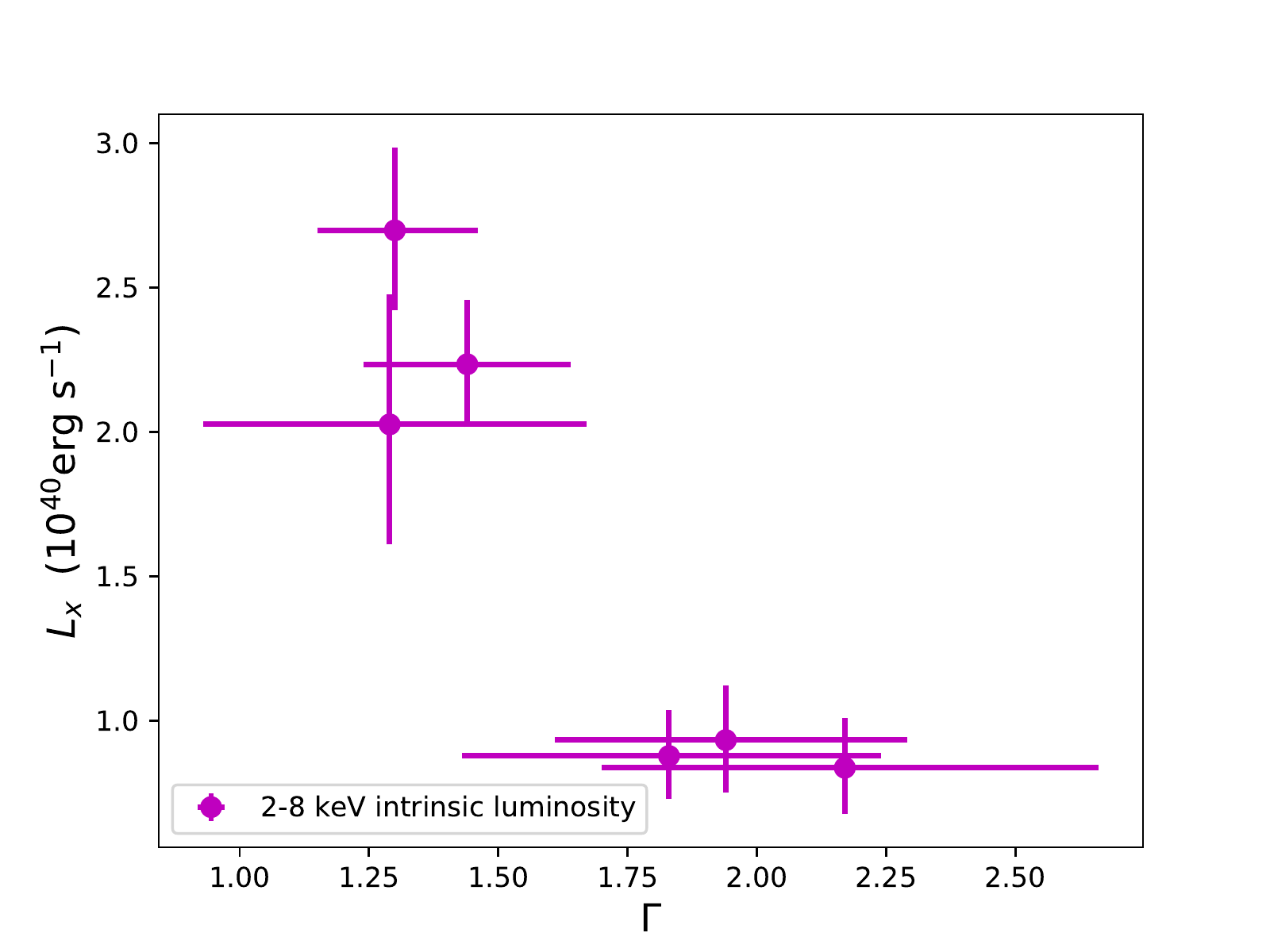}
    \caption{$\Gamma-\rm L_X$ negative correlation for different \xmm\ epochs for $0.3-8.0$ keV energy range (Left). Same quantities are plotted on the right side also but for 2.0-8.0 keV energy range. }
    \label{fig:anticorr}
\end{figure*}

NGC 1042 ULX1 shows a negative correlation between spectral photon index and luminosity (see figure \ref{fig:anticorr} and section \ref{sec:anti-corr}), which means a higher luminosity state is harder. One of the possible explanations for such a behavior is that the hard photons are aligned to the line of sight through the optically thin tunnel, and with a higher accretion rate (which means higher luminosity) in the inner region of the accretion disk, hot (hard) photons can reach us. This would imply that, in general, NGC 1042 ULX1 is a low inclination system where the outflow is optically thin. Hence, an increase in accretion rate does not ensure that the hard photons would be down-scattered and move out of the line of sight. Moreover, geometrical beaming would play a crucial role to explain that the higher luminosity state is spectrally harder. The hard emission which originates from the inner accretion flow is beamed towards the line of sight through the optically thin tunnel, hence the hard emission would be more intensified compared to the softer component with increasing accretion rate (e.g., \citealt{Poutanen2007, King2009, Middleton2015, Luangtip2016}).

While discussing the Comptonization process, one should consider both black hole and neutron star scenarios of ULXs. This Comptonization can be either an external Comptonization in the Corona region due to inverse-scattering, or it can be a magnetized Comptonization due to shock formation in the polar region of a neutron star. The hard sources are often perceived to be strongly magnetized neutron star systems where the emission is directly coming from the accretion column \citep{Gurpide2021a}.

\subsection{Nature of the accretor}
The acceleration search in the Fourier space of time series does not confirm any pulsation candidates, hence we cannot confirm whether ULX1 hosts a neutron star or black hole. An interesting comparison of NGC 1042 ULX1 and its hard spectra would be with the spectra of pulsar ULXs and the sources which resemble pulsar-like spectra (see, e.g., \citealt{Pintore2017, Walton2018}). The apparent missing spectral curvature in \xmm\ data is clearly seen in the new \nustar\ data, recognizing that this source has the spectral curvature like many other ULXs in the super-Eddington accretion state. From our analysis, we see the presence of cool accetion disk component with a characteristic temperature similar to the ultraluminous state sources described in \citealt{Sutton2013}. Moreover, we find the presence of characteristic spectral curvature in ULX1 like other ULX sources. Based on these findings, we can discard the possibility of ULX1 being a low/hard state sub-Eddington IMBH system. Within the limitation of currently available data, the spectral similarities with ultraluminous state sources indicate that ULX1 is a super-Eddington accretor (either a stellar-mass black hole or a neutron star).

\subsection{Anti-correlation between $\Gamma-\rm L_X$} \label{sec:anti-corr}
The correlation between $\Gamma$ and $\rm L_X$ are widely studied for X-ray binaries (e.g., \citealt{Yang2015}) and ULXs (e.g., \citealt{Kajava2009}). They are excellent probes in understanding the physical accretion processes in these sources. We carry out a similar study for NGC 1042 ULX1 and plot the correlation between luminosity and spectral photon index in figure \ref{fig:anticorr}. We discuss the theoretical notion of negative correlation found for ULX1. Before that, we need to discuss possible ``artefacts" which can arise from absorption and low-energy thermal components.

We find anti-correlation between $\Gamma$ and $\rm L_X$ for \texttt{powerlaw} fit in both cases when the absorption parameter is kept free for all epochs and linked between different epochs. In both cases, unabsorbed and absorbed luminosities are negatively correlated with $\Gamma$. The overall nature of this negative correlation is not significantly influenced by the absorption parameter, as shown in figure \ref{fig:free_nh_parameters}-right. However, a reader would be cautious when interpreting the correlation between $\Gamma$ and $\rm N_H$, where a slightly lower $\rm N_H$ trend is seen when the spectra are harder. The Pearson r-coefficient \footnote{\url{https://docs.scipy.org/doc/scipy/reference/generated/scipy.stats.pearsonr.html}} is $\sim 0.77$ with p-value $\sim 0.04$ and the Spearman correlation coefficient \footnote{\url{https://docs.scipy.org/doc/scipy/reference/generated/scipy.stats.spearmanr.html}} is $\sim 0.73$ with p-value $\sim 0.06$. The p-values suggest the probability of having the same correlation measurement from an uncorrelated system. In other words, typically, if $\rm p>0.05$, then the correlation might have occurred by chance and cannot be considered statistically significant. Although this correlation between $\rm N_H$ and $\Gamma$ is not statistically significant, we would discuss two possibilities for such a trend. First, the \texttt{powerlaw} model extends to lower energy arbitrarily, degenerating with softer spectral components. This could give rise to such a correlation. Another possibility is that the $\rm N_H-\Gamma$ correlation is indeed physical. A higher accretion rate would increase the luminosity and we see a trend of increasing hardness in the system. Due to low inclination, the hard photons pass through the optically thin tunnel and reach us. Hence, the neutral $\rm N_H$ component appears to be less dominant with the harder spectral state. In either case, the anti-correlation between $\Gamma$ and $\rm L_X$ is prominent and can be understood as real. When the absorption parameter is linked, we measure the correlation coefficients between $\Gamma$ and $\rm L_X$. For absorbed luminosity, the Pearson r-coefficient is $\sim -0.921$ with p-value $\sim 0.003$ and the Spearman correlation coefficient is $\sim -0.857$ with p-value $\sim 0.014$. For unabsorbed luminosity, the Pearson r-coefficient is $\sim -0.857$ with p-value is $\sim 0.014$ and the Spearman correlation coefficient is $\sim -0.857$ with p-value $\sim 0.014$ (see figure \ref{fig:anticorr} - left). 

To remove any ``artefact" from soft energy regime apart from the neutral absorption, we also study the correlation trend between $2.0-8.0$ keV $\Gamma$ and intrinsic $\rm L_X$ (figure \ref{fig:anticorr} - right). This would minimize any artificial boost in luminosity in the soft energy part and predict the correlation only in the high energy spectrum. For this purpose, we fit the data with an absorbed \texttt{powerlaw} model in the $2.0-8.0$ keV range by fixing the $\rm N_H$ to the best-fit value from $0.3-8.0$ keV fit. The negative correlation still holds although the measurement uncertainties are large due to lower count statistics. Observation XM1 has very low counts in $2.0-8.0$ keV; hence, the measurement errors are very high. Hence, we do not consider this observation in figure \ref{fig:anticorr} - right. In this figure, the Pearson r-coefficient is $\sim -0.922$ and p-value is $\sim 0.009$. The Spearman correlation coefficient, on the other hand, for the same data points, is $\sim -0.771$, and the p-value is $\sim 0.072$. We would also mention that the measurement of the Pearson and Spearman correlation coefficients need to be treated with caution because of the low sample space and also these measurements do not consider the errors in the parameters.

It is important to understand the underlying theoretical interpretation for the appearance of a such anti-correlation property. We compare with the results and interpretation of \citet{Yang2015} in the context of XRBs in general. Although the systems studied in \citet{Yang2015} have very low Eddington ratio ($\frac{\rm L_X}{\rm L_{Edd}}$ varying between $\sim 10^{-8.5}$ to $10^{-1.5}$), and warrant a cautious comparison while explaining similar properties for super-Eddington sources like ULXs, the interpretation of the underlying physics is interesting and can be formally explored in the context of ULXs in future studies. The explanation for the negative correlation is attributed to the Type I luminous hot accretion flow (LHAF) in systems where the accretion rate is high but not too high that the accretion time scale is shorter than the growth time scale of thermal instability. The electrons in the plasma would be radiatively efficient unless the density is too low, which means total luminosity can be approximated to hard luminosity ($\rm L_{hard}$). During a higher accretion rate, the synchrotron absorption depth would increase, and soft luminosity ($\rm L_{soft}$) will not sufficiently increase, which leads to the negative correlation between $\Gamma$ and $\rm L_X$. As mentioned before, comparison with this interpretation, needs to be cautioned since, NGC 1042 ULX1 is possibly a stellar mass system, as described above. Hence, the super-Eddington interpretation would be a much more viable scenario in this case. Nevertheless, increasing synchrotron absorption depth with increasing accretion rate could still be a physical mechanism that can explain the negative $\Gamma-\rm L_X$ correlation.

The negative correlation between $\Gamma$ and $\rm L_X$ can also be interpreted as a higher Compton up-scattering photon fraction due to a higher accretion rate (i.e., higher luminosity). For example, in a pulsar system, when the mass accretion rate increases, the pressure, and electron density increase, which eventually leads to a higher optical depth and higher $y = \frac{\tau k T_e}{m_e c^2}$ parameter in Comptonization process. Thus the Compton up-scattered photon manifests as a harder spectrum due to a higher accretion rate (see, e.g., \citealt{Malacaria2015} and references therein). Although, as cautioned by \citealt{Malacaria2015}, this interpretation is primarily valid for sub-critical sources, we argue that similar phenomena are feasible for super-critical sources also, since the broadband study of many super-Eddington ULX sources (e.g., \citealt{Kaaret2017, Walton2020}) predicts a strong contribution of Compton scattered photons in the hard spectral regime.

In the direct context of modern understanding of ULX spectral properties, this harder when brighter correlation can be interpreted as geometric beaming of hard photons directly to the line of sight as explained in section \ref{sec:accretion_state}. Also, the hard spectrum is often explained by the emission from the strong accretion column (e.g., \citealt{Pintore2017, Walton2018, Gurpide2021a}). Relative contribution of change in accretion rate and magnetic field strength would determine the scale of magnetospheric radius $\rm R_M$ and spherization radius $\rm R_S$. If $\rm R_M$ truncates the disk close to the $\rm R_S$, the spectral contribution from accretion column would be stronger (see \citealt{Walton2018}), hence making the spectra harder. That would mean the increasing luminosity is directly coming from the hard emission from the accretion column. This is also supported by the observed trend in table \ref{tab:tablediskpow}, where harder spectral epochs like XM2 and XM4 are seen to have increased flux in powerlaw component.

To summarize, NGC 1042 ULX1 is a bright ULX whose luminosity reaches a few times $\sim 10^{40}$ \lumcgs. The spectral properties suggest that the source is a stellar mass super-Eddington accretor. Due to variations in accretion rate, disk occultation, or strength of Compton scattering, the source exhibit a change in luminosity and spectral hardness. Future monitoring broadband observations can decipher the energy-dependent variability in the source beyond $\sim 10$ keV and help in understanding the accretion state of the source in a more comprehensive form.

\section*{Acknowledgements}
We would like to thank the referee for the useful suggestions which have helped to further improve the manuscript. We would like to thank M. Bachetti for the helpful discussions at various stages of the work. This research has utilized data obtained with \nustar, a project led by Caltech, funded by NASA, and managed by the NASA Jet Propulsion Laboratory (JPL), and has used the \texttt{NuSTARDAS} software package, jointly developed by the Space Science Data Centre (SSDC; Italy) and Caltech (USA). This research has also utilized data obtained with \xmm, an ESA science mission with instruments and contributions directly funded by the ESA Member States and NASA.

%The Acknowledgements section is not numbered. Here you can thank helpful colleagues, acknowledge funding agencies, telescopes and facilities used etc. Try to keep it short.

%%%%%%%%%%%%%%%%%%%%%%%%%%%%%%%%%%%%%%%%%%%%%%%%%%
\section*{Data Availability}
The \xmm\ data utilized in this work are available for download in the High Energy Astrophysics Science Archive Research Center (HEASARC) archive (\url{https://heasarc.gsfc.nasa.gov/db-perl/W3Browse/w3browse.pl}). The \nustar\ data used in this work were acquired from the joint \nicer+\nustar\ proposal and will become available in the HEASARC archive from December 2022.
 
%The inclusion of a Data Availability Statement is a requirement for articles published in MNRAS. Data Availability Statements provide a standardised format for readers to understand the availability of data underlying the research results described in the article. The statement may refer to original data generated in the course of the study or to third-party data analysed in the article. The statement should describe and provide means of access, where possible, by linking to the data or providing the required accession numbers for the relevant databases or DOIs.



\bibliographystyle{mnras}
\bibliography{NGC1042} % if your bibtex file is called example.bib

%\begin{thebibliography}{99}
%\end{thebibliography}

%%%%%%%%%%%%%%%%%%%%%%%%%%%%%%%%%%%%%%%%%%%%%%%%%%

%%%%%%%%%%%%%%%%% APPENDICES %%%%%%%%%%%%%%%%%%%%%

%\appendix



%%%%%%%%%%%%%%%%%%%%%%%%%%%%%%%%%%%%%%%%%%%%%%%%%%

% Don't change these lines
\bsp	% typesetting comment
\label{lastpage}
\end{document}